\documentclass[aps,prl,twocolumn,superscriptaddress,amsmath,amssymb]{revtex4}

\usepackage{graphicx}

\begin{document}

\title{Rheology of fractal networks}

\author{P. Patr\'{\i}cio}
\email{pedro.patricio@adf.isel.pt}
\affiliation{ISEL - Instituto Superior de Engenharia de Lisboa, Instituto Polit\'ecnico de Lisboa, 1959-007 Lisboa, Portugal.}
\affiliation{CEDOC, Faculdade de Ci\^encias M\'edicas, Universidade Nova de Lisboa, 1169-056 Lisboa, Portugal.}

\author{C. R. Leal}
\affiliation{ISEL - Instituto Superior de Engenharia de Lisboa, Instituto Polit\'ecnico de Lisboa, 1959-007 Lisboa, Portugal.}
\affiliation{Centro de Investiga\c c\~ao em Agronomia, Alimentos, Ambiente e Paisagem, LEAF,\\ 
Instituto Superior de Agronomia, Universidade de Lisboa, 1349-017 Lisboa, Portugal.}

\author{J. Duarte}
\affiliation{ISEL - Instituto Superior de Engenharia de Lisboa, Instituto Polit\'ecnico de Lisboa, 1959-007 Lisboa, Portugal.}
\affiliation{CAMGSD, Instituto Superior T\'ecnico, Universidade de Lisboa, 1049-001 Lisboa, Portugal}

\author{C. Janu\'ario}
\affiliation{ISEL - Instituto Superior de Engenharia de Lisboa, Instituto Polit\'ecnico de Lisboa, 1959-007 Lisboa, Portugal.}

\date{July 20, 2015}

\begin{abstract}

We model the cytoskeleton as a fractal network by identifying each segment with a simple Kelvin-Voigt element, with a well defined equilibrium length.
The final structure retains the elastic characteristics of a solid or a gel, which may support stress, without relaxing.
By considering a very simple regular self-similar structure of segments in series and in parallel, in 1, 2 or 3 dimensions, 
we are able to express the viscoelasticity of the network as an effective generalised Kelvin-Voigt model with a power law spectrum of retardation times, 
$\cal L\sim\tau^{\alpha}$. We relate the parameter $\alpha$ with the fractal dimension of the gel. 
In some regimes ($0<\alpha<1$), we recover the weak power law behaviours
of the elastic and viscous moduli with the angular frequencies, $G'\sim G''\sim w^\alpha$, that occur in a variety of soft materials, including living cells.
In other regimes, we find different power laws for $G'$ and $G''$.

\end{abstract}

\maketitle

Microrheology measurements on the cytoskeleton of the cell revealed interesting
weak power law behaviours \cite{fabry2001scaling} (see \cite{kollmannsberger2011linear,pritchard2014mechanics} for recent reviews), 
which are frequently associated to the phenomenological ``Soft Glassy Materials'' (SGM) model \cite{sollich1997rheology,sollich1998rheological}.
Based on the idea of structural disorder and metastability, common to all SGMs, this model relates the power law exponent
of the elastic and viscous moduli, $G'(w)\sim G''(w)\sim w^\alpha$, with $0<\alpha<1$,
to a mean-field noise temperature $x=\alpha+1$, with a glass transition occurring at $x=1$.

However, rather than a generic fluidic system above the glass transition, 
the cytoskeleton could be more easily associated with a polymer network near the sol-gel transition.
Using the ideas of percolation and self-similarity, 
the power law exponent $\alpha$ has been previously related to the fractal dimension $d_f$ of a flexible polymer cluster.
This relationship is not unique and depends on the underlying assumptions of the proposed microscopic models (monodispersity vs polydispersity, 
unscreening vs screening of excluded volume, etc. -- see \cite{winter1997rheology} for a review).
In particular, if a polydisperse polymeric fractal (prescribed by bond percolation theory), 
following Rouse chain dynamics (for flexible polymer chains) with fully screened hydrodynamic interactions is considered \cite{muthukumar1989screening}, 
the exponent $\alpha$ can take values between 0 and 1 for $d_f$ ranging from 2.5 to 1.25, respectively.

More recently, the weak power law behaviour of the cytoskeleton has been associated with
the ``Glassy Worm Like Chain" (GWLC) model \cite{kroy2007glassy} (so-called from its analogy to SGM model).
This model defines an average separation between the crosslinks along the filaments.
If the relaxation modes have a wavelength shorter than this separation, its relaxation time follows the ``Worm Like Chain" (WLC) model for semiflexible polymers, 
which largely compose the cytoskeleton.
Otherwise, the relaxation spectrum is stretched through an effective Boltzmann factor with a characteristic energy $\epsilon$ that must be overcome 
to induce a conformational change of the network. 
This model predicts a high frequency regime with a power law exponent $\alpha=3/4$ (corresponding to WLC model) and a low frequency regime with a second power law exponent that depends on the phenomenological parameter $\epsilon$. 
Extensions of this model deal with the possibility of transient crosslinking between filaments \cite{wolff2010inelastic}.

It is known that the cytoskeleton, or a cellular tissue, even in equilibrium, supports a certain amount of stress, which is imposed
by a substrate or other neighbouring cells. The structural cytoskeleton filaments must retain a solid character, without a full relaxation.
In this article, we will consider a very simple model of a solid gel,  
composed of a regular self-similar network of segments with well defined lengths and rigid bonds. 
By identifying each segment with a Kelvin-Voigt viscoelastic element (an hypothesis suggested in 
\cite{balland2006power,patricio2015weak}), we are able 
to express the viscoelasticity of the network as an effective generalised Kelvin-Voigt model with a power law spectrum of retardation times, 
$\cal L\sim\tau^{\alpha}$, where $\alpha$ is related to the network power law distribution of lengths (and eventually to its fractal dimension),
and to the Kelvin-Voigt particular characteristics of each segment.
This relation is not direct, since in 2 or 3 dimensions we have a large collection of Kelvin-Voigt elements in series and in parallel.
When $0<\alpha<1$, we recover the weak power law behaviours $G'\sim G''\sim w^\alpha$. 
In other regimes, for $\alpha<0$ and $\alpha>1$, we obtain, first analytically (with some approximations), and then numerically,
different and interesting power law behaviours for $G'$ and $G''$. 

\begin{figure}[h]
\begin{center}
\includegraphics[scale=0.5]{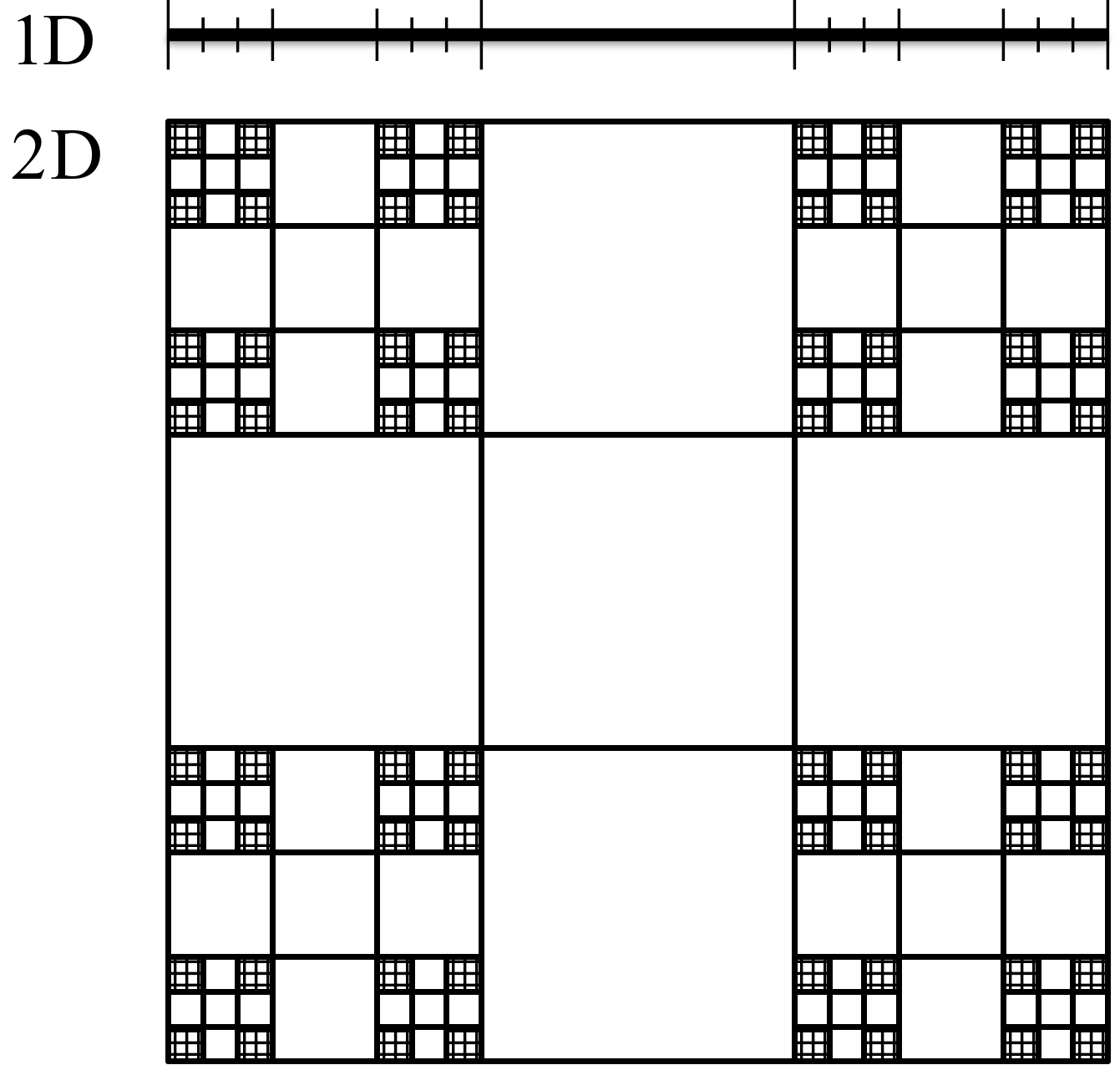}
\caption{Top: simple 1D self similar structure of segments. Bottom: 2D network of segments generated from the 1D self similar structure represented at the top.}
\label{fig 1} 
\end{center}
\end{figure}

Let us initially consider a simple 1D self-similar structure that is defined 
by first dividing the system size $L$ by a number $N_L$.
We will get $N_L$ segments of size $L/N_L$. 
Now, we pick only $K_L<N_L$ segments and repeat the procedure $n$ times
(see Fig. \ref{fig 1}, top, for $N_L=3$ and $K_L=2$).
We obtain $(N_L-K_L)K_L^{i-1}$ segments of size $l_i=L/N_L^i$ (with $1<i<n-1$), and $N_L K_L^{n-1}$ segments of size $l_n=L/N_L^n$.

We may associate each segment, of size $l$, with a Kelvin-Voigt element, composed 
of a spring of stiffness $G_l$ in parallel with a dashpot of viscosity $\eta_l$. 
The (undeformed) spring ensures the length $l$. The element behaves elastically on long times scales and its dynamics comes from the viscous element.
Its creep compliance is $J_l(t)=J_l(1-e^{-t/\tau_l})$ \cite{phan2012understanding}, where $J_l=1/G_l$ and $\tau_l=\eta_l/G_l$ is the element retardation time. 
Because the elements are in series, the total creep compliance is just the sum of all the elements' creep compliances \cite{phan2012understanding}. 
Taking into account the repetition of segment sizes, which are a consequence of the self similar construction, we may write
\begin{equation}
J(t)=(N_L-K_L)\sum_{i=1}^{n-1}K_L^{i-1}J_{l_i}(t)+N_L K_L^{n-1}J_{l_n}(t)
\label{eq 1D}
\end{equation} 

This sum may be approximated by an integral by multiplying it by $di=1$. 
If $di$ is considered to be small, we may write
\begin{equation}
J(t)\sim \int_{l_1}^{l_n}K_L^i J_{l}(t)\frac{di}{dl}dl
\end{equation}
The product $|K_L^idi/dl|$ corresponds to the density of segments of size $l$ (per unit size).
If we invert the relation $l=l_i=L/N_L^i$, we obtain
\begin{equation}
i=\frac{1}{\ln N_L}\ln\frac{L}{l},\;\;\;\;\;\frac{di}{dl}=-\frac{1}{\ln N_L}\frac{1}{l}
\end{equation}
and we may write
\begin{equation}
K_L^i=e^{i\ln K_L}=\left(\frac{L}{l}\right)^{\xi},\;\;\;\;\;\xi=\frac{\ln K_L}{\ln N_L}
\end{equation}
Thus, the total creep compliance becomes
\begin{equation}
J(t)\sim \int_{l_n}^{l_1} l^{-\xi-1} J_{l}(t)dl
\label{eq 1D int}
\end{equation}
where $l_{min}=l_n$ and $l_{max}=l_1$.

Let us now suppose, in very general terms, that $J_l\sim l^\delta$ and $\tau=\tau_l\sim l^\gamma$, with $\gamma>0$. 
In these conditions, we have
\begin{align}
J(t)&\sim \int_{\tau_n}^{\tau_1} l^{-\xi-1} J_l \left(1-e^{-t/\tau_l}\right) \frac{dl}{d\tau}d\tau\nonumber\\
&\sim \int_{\tau_n}^{\tau_1} \tau^{(\delta-\xi)/\gamma}\left(1-e^{-t/\tau}\right)\frac{d\tau}{\tau}
\label{eq creep int}
\end{align}
where $\tau_{min}=\tau_n$ and $\tau_{max}=\tau_1$.

The function defined by $\cal L \sim  \tau^{(\delta-\xi)/\gamma}$ 
corresponds to the generalised Kelvin-Voigt retardation spectrum \cite{phan2012understanding}.
If we make the association
\begin{equation}
\alpha^{1\text D}=\alpha=\frac{\delta-\xi}{\gamma}
\end{equation}
we may write $\cal L \sim  \tau^{\alpha}$. The rheological response of the structure depends essentially on this parameter $\alpha$,
and the minimum and maximum retardation times. 

To calculate the elastic and viscous moduli, we determine first the Laplace transform of the creep compliance:
\begin{align}
\tilde J(s) &= \int_0^\infty J(t)e^{-st}ds
\sim \int_{\tau_n}^{\tau_1} \tau^{\alpha-1}\frac{1}{s(1+s\tau)}d\tau
\label{eq creep lap}
\end{align}
Then, the complex creep compliance, which is given by the relation $J^*(w)=(iw)\tilde J(iw)$.
The elastic and viscous compliances, defined through $J^*=J'-iJ''$, are respectively given by
\begin{align}
\label{eq elastic compliance}
J'(w)&\sim \int_{\tau_n}^{\tau_1} \tau^{\alpha-1}\frac{1}{1+(w\tau)^2}d\tau \\
J''(w)&\sim \int_{\tau_n}^{\tau_1} \tau^{\alpha-1}\frac{w\tau}{1+(w\tau)^2}d\tau
\label{eq viscous compliance}
\end{align}
Finally, the complex modulus is related to the complex compliance through $G^*J^*=1$.
The elastic and viscous moduli, defined by $G^*=G'+iG''$, may be determined from the relations:
\begin{align}
G'=\frac{J'}{J'^2+J''^2},\;\;\;
G''=\frac{J''}{J'^2+J''^2}
\end{align}

For $w \ll 1/\tau_1$ ($w\tau\ll 1$ for any $\tau<\tau_{max}=\tau_1$), the integrals of Eq. \ref{eq elastic compliance} and \ref{eq viscous compliance}
are much simplified, and we obtain the scalings $J'\sim w^0$ and $J''\sim w^1$, with $J'\gg J''$. The elastic and viscous moduli
scale as
\begin{align}
G'\approx\frac{1}{J'}\sim w^0,\;\;\;G''\approx\frac{J''}{J'^2}\sim w^1
\end{align}

For $w \gg 1/\tau_n$ ($w\tau\gg 1$ for any $\tau>\tau_{min}=\tau_n$), the integrals of Eq. \ref{eq elastic compliance} and \ref{eq viscous compliance}
are again simplified, yielding the scalings $J'\sim w^{-2}$ and $J''\sim w^{-1}$, with $J'\ll J''$. 
We obtain then the same scalings for the elastic and viscous moduli:
\begin{align}
G'\approx\frac{J'}{J''^2}\sim w^0,\;\;\;G''\approx\frac{1}{J''}\sim w^1
\end{align}

The coefficients of $G'$ and $G''$ are dependent of $\alpha$ but their scalings are not. In fact, the exponents of $w$ 
coincide with the scalings of a simple Kelvin-Voigt model.

For $1/\tau_1\ll w \ll 1/\tau_n$, we may in some cases extend the limits of the integrals of Eq. \ref{eq elastic compliance} and \ref{eq viscous compliance}
to $\tau_{min}\to 0$ and $\tau_{max}\to\infty$, which allow us to obtain the results:
\begin{align}
\label{eq elastic compliance 2}
J'(w)&\sim \frac{\pi}{2} w^{-\alpha}\text{csc}\frac{\pi \alpha}{2}\;\;\;(\text{if}\;\;0<\alpha<2) \\
J''(w)&\sim \frac{\pi}{2} w^{-\alpha}\text{sec}\frac{\pi \alpha}{2}\;\;\;(\text{if}\;\;-1<\alpha<1) 
\label{eq viscous compliance 2}
\end{align}
When $0<\alpha<1$, both integrals are well defined and we recover the weak power law behaviours:
\begin{equation}
G'(w)\sim G''(w)\sim w^\alpha,\;\;\;\;\frac{G''}{G'}=\tan\frac{\pi \alpha}{2}
\label{eq SGM behaviour}
\end{equation}
If $\alpha<0$, the integral for the elastic compliance (Eq. \ref{eq elastic compliance}) diverges as $\tau_{min}=\tau_n\to 0$.
The elastic compliance is then dominated by the smallest retardation times. 
In this case, we have $w\tau_n\ll 1$, and $J'\sim w^0$.
On the contrary, if $\alpha>2$, then this integral is dominated by the largest retardation times, $\tau_{max}=\tau_1$. 
We have $w\tau_1\gg 1$, and the power law behaviour $J'\sim w^{-2}$.
By the same line of reasoning, we may determine from Eq. \ref{eq viscous compliance} 
the power law behaviours $J''\sim w^1$ for $\alpha<-1$ and $J''\sim w^{-1}$ for $\alpha>1$.
Applying these results to each interval of $\alpha$, and using the approximations $J'\gg J''$ for $\alpha<0$ and $J'\ll J''$ for $\alpha>1$
(which we may infer from Eq. \ref{eq elastic compliance 2} and \ref{eq viscous compliance 2}), we obtain the power law behaviours:
\begin{align}
&G'\sim w^0 & &G''\sim w^1 & &(\alpha<-1) \\
&G'\sim w^0 & &G''\sim w^{-\alpha} & &(-1<\alpha<0) \\
&G'\sim w^\alpha & &G''\sim w^{\alpha} & &(0<\alpha<1) \\
&G'\sim w^{1-\alpha} & &G''\sim w^1 & &(1<\alpha<2) \\
&G'\sim w^0 & &G''\sim w^1 & &(2<\alpha)
\end{align}
We note that for $\alpha<-1$ or $\alpha>2$, we have a single power law behaviour for $G'\sim w^0$ and $G''\sim w^1$, for all values of the angular frequency $w$.
Indeed, for these ranges of $\alpha$, the whole structure is entirely dominated by only one Kelvin-Voigt element, 
corresponding respectively to the minimum ($\alpha<-1$) or the maximum ($\alpha>2$) retardation times.

\begin{figure}[h]
\begin{center}
\includegraphics[scale=0.35]{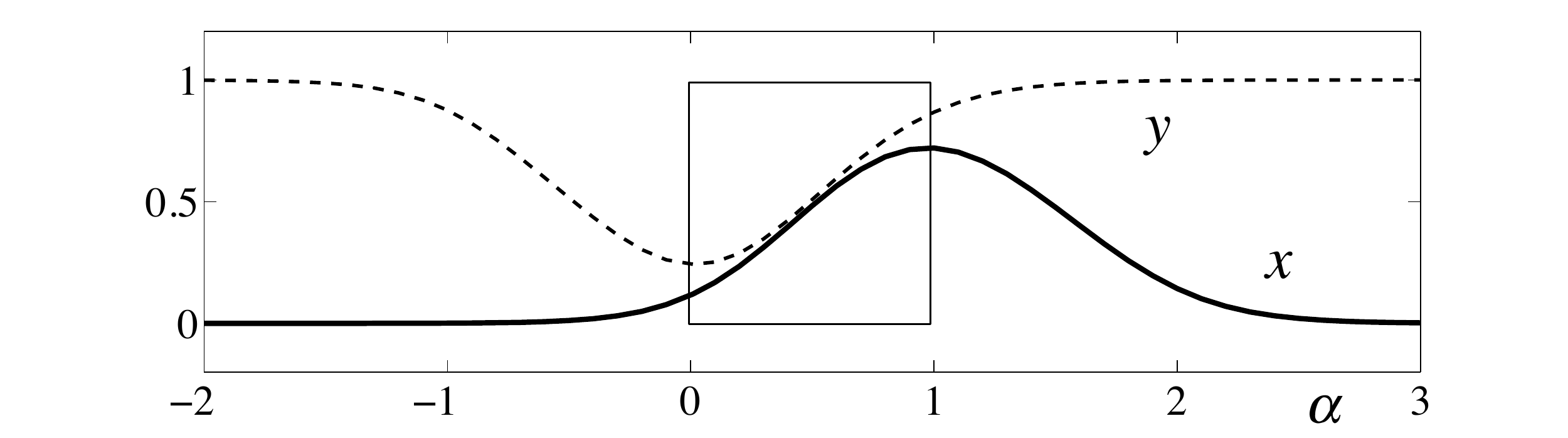}
\caption{Exponents $x$ and $y$ vs $\alpha$ of the power laws $G'\sim w^x$ and $G''\sim w^y$,
for a generalised Kelvin-Voigt model with a power law spectrum of retardation times $\cal L\sim\tau^{\alpha}$, for $\tau_{max}/\tau_{min}=10^{6}$.
The highlighted region refers to the weak power law behaviour, in which $x\approx y\approx \alpha$ ($0<\alpha<1$).}
\label{fig 2} 
\end{center}
\end{figure}

\begin{figure}[htp]
\begin{center}
\begin{tabular}{cc}
\includegraphics[scale=0.17]{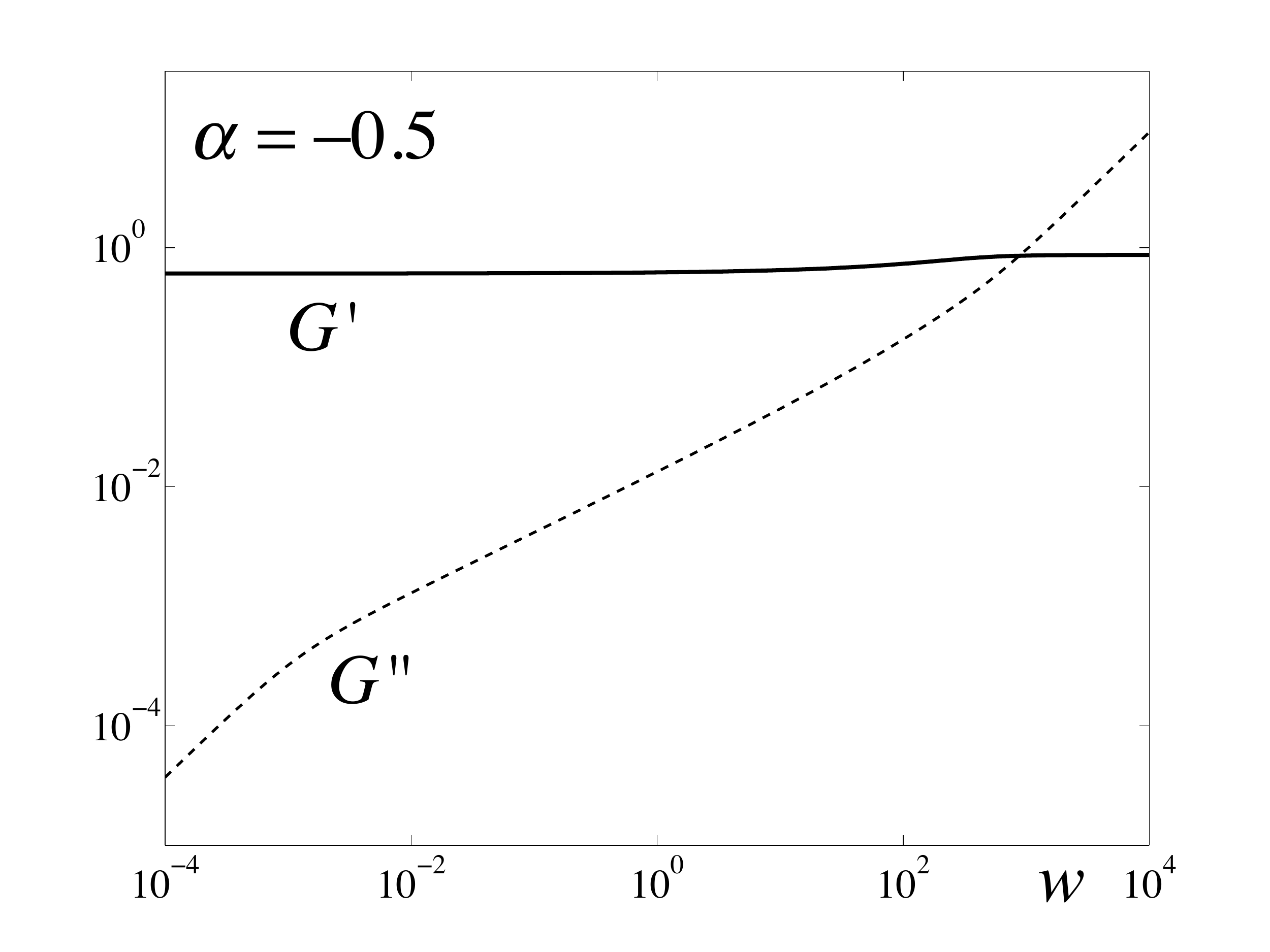}\includegraphics[scale=0.17]{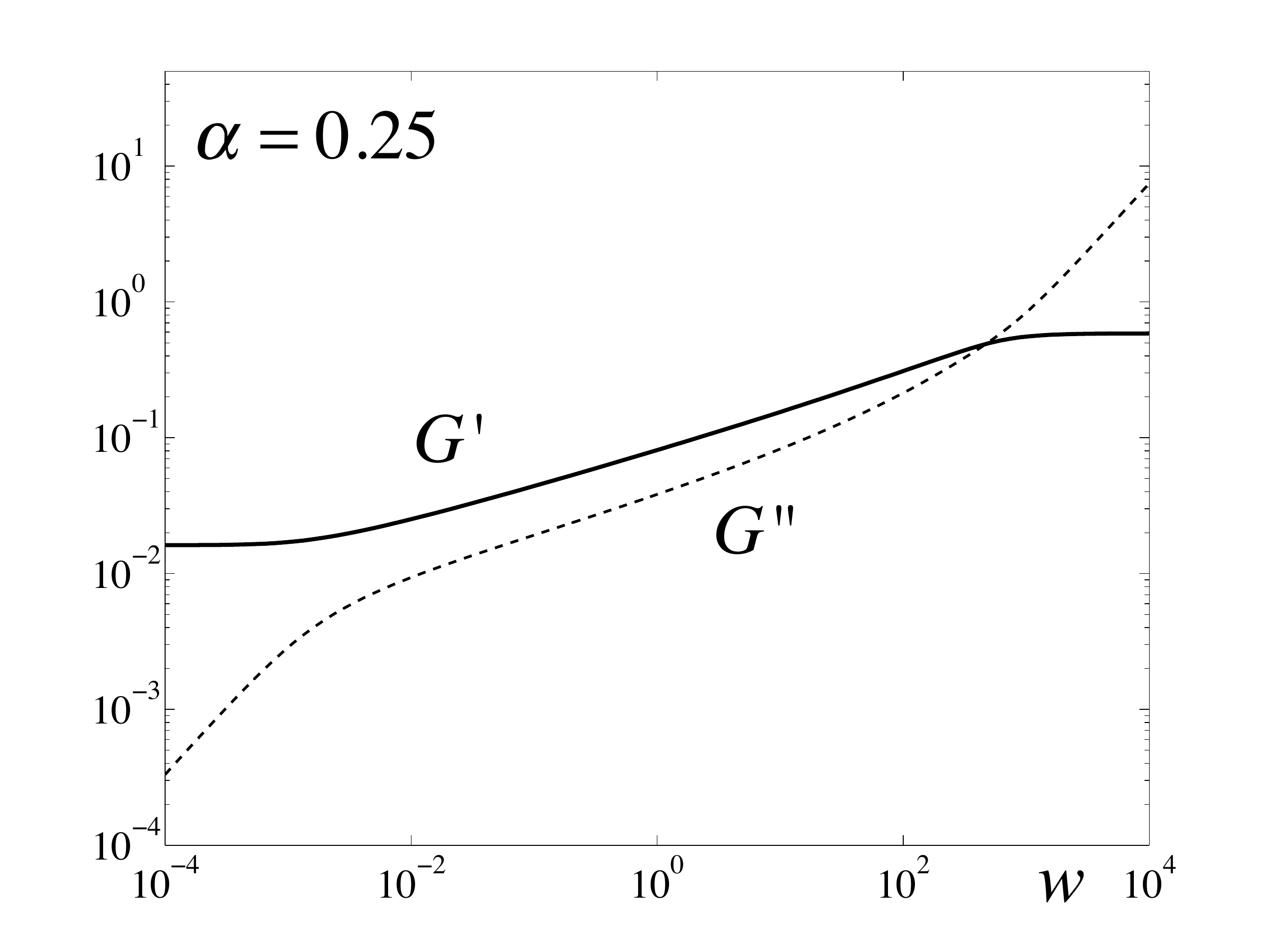}\\
\includegraphics[scale=0.17]{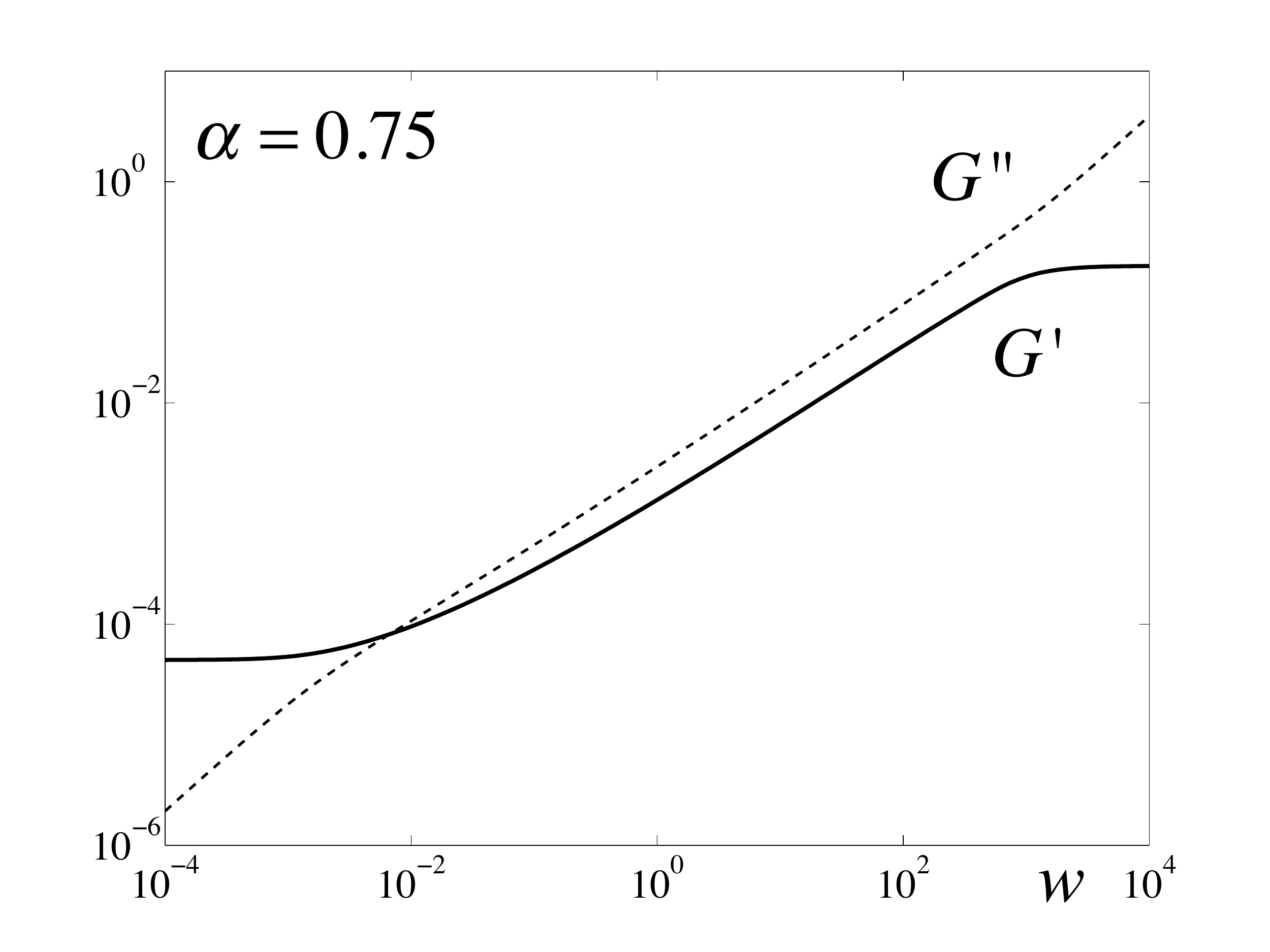}\includegraphics[scale=0.17]{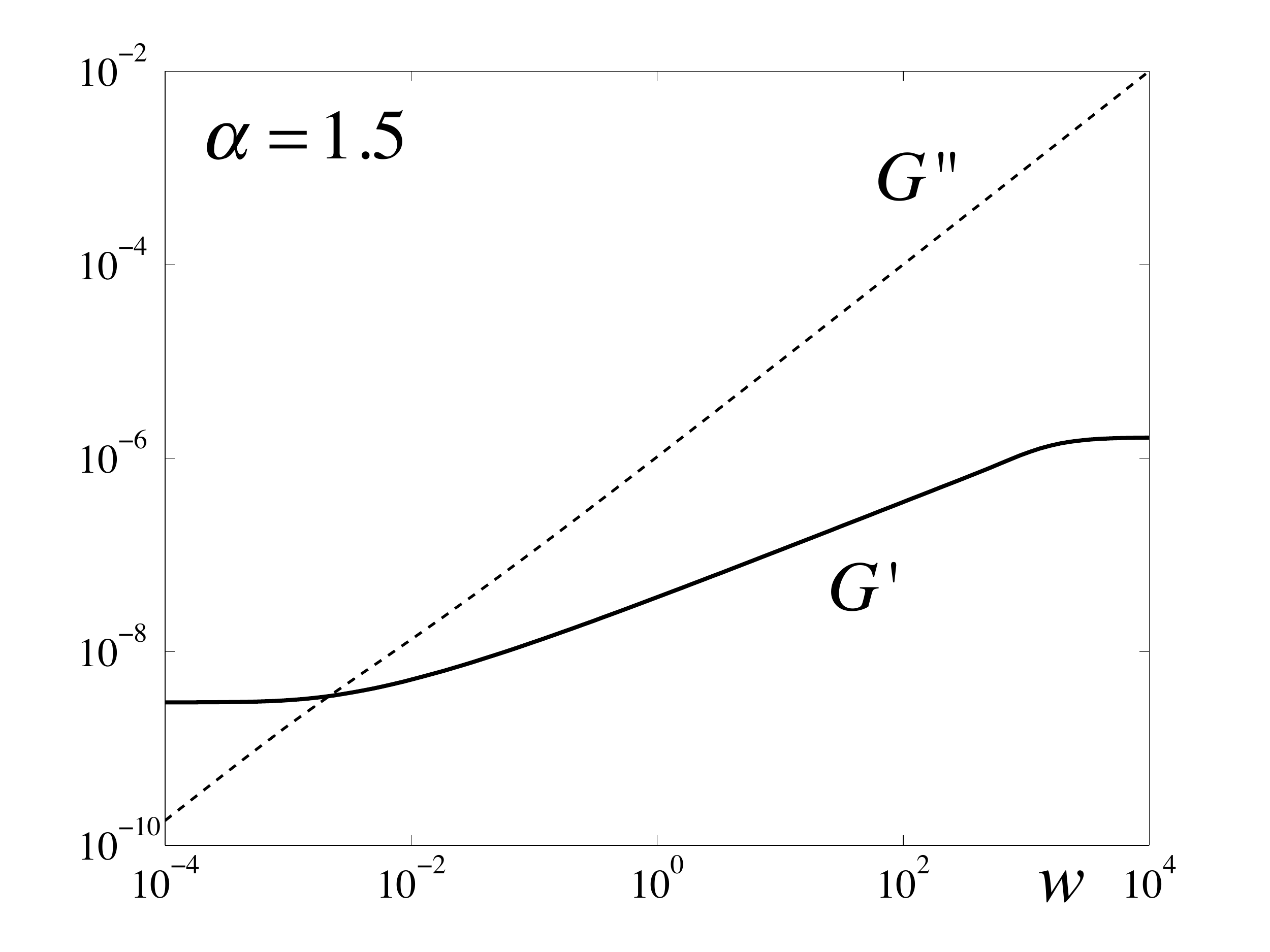}
\end{tabular}
\caption{Elastic and viscous moduli, $G'(w)$ (solid line) and $G''(w)$ (dashed line)
for a generalised Kelvin-Voigt model with a power law spectrum of retardation times $\cal L\sim\tau^{\alpha}$,
with $\alpha=-0.5,0.25,0.75,1.5$, $\tau_{min}=10^{-3}$, $\tau_{max}=10^3$ (in arbitrary units).}
\label{fig 3}
\end{center}
\end{figure}

The creep compliance of the 1D self similar structure (Eq. \ref{eq 1D}) corresponds to an effective discrete generalised Kelvin-Voigt model.
The accurate values for the exponents of the power law behaviours, $G'\sim w^x$ and $G''\sim w^y$,
for any value of $\alpha$, are shown in Fig. \ref{fig 2}, for $\tau_{max}/\tau_{min}=10^6$.
Representative plots of $G'(w)$ and $G''(w)$ are shown in Fig. \ref{fig 3} (see \cite{patricio2015weak} for calculation details).

To create a 2D or a 3D network structure of segments, we may follow the same self similar construction in the other dimensions 
(see 2D network in Fig. \ref{fig 1}, bottom).
The boundaries between the segments are identified with the network crosslinks, which allow us to extend the segments
into the interior part of the system, of size $L^2$ or $L^3$.

The total creep compliance reflects the deformation of the 2D or 3D network along the direction of the applied force, or stress, 
which we take to be the direction of the 1D structure considered before.
But now we have an intricate combination of segments in series and in parallel. 

Let us consider first the 2D network structure depicted in Fig. \ref{fig 1}, bottom. 
In the first iteration of our construction, we have 
the boundary segments of the square of size $L$ divided in $N_L=3$ parts. 
At the boundaries between the segments, we extend new segments into the interior of the system, creating
a matrix of $N_L\times N_L$ adjacent squares of size $L/N_L$. We have thus generated $M_L=N_L+1$ lines in the direction of the force.
These lines are in parallel. Each line is composed of $N_L$ segments of size $L/N_L$, in series. 
This regular structure of $M_L$ lines in parallel with $N_L$ segments in series, each of one with equal creep compliance, gives the matrix creep compliance:
\begin{equation}
J^{\text{mat}}_{l_1}(t)=\frac{N_L}{M_L}J_{l_1}(t)
\label{eq 2D matrix}
\end{equation}
At the $n^\text{th}$ iteration, after considering all the different elements in series and in parallel, we get a surprisingly simple result.
In fact, the number of segments in series cancels with the number of segments in parallel, yielding:
\begin{equation}
J^{2\text{D}}(t)=\frac{N_L-K_L}{N_L+1}\sum_{i=1}^{n-1}J_{l_i}(t)+\frac{N_L}{N_L+1}J_{l_n}(t)
\label{eq 2D}
\end{equation}

In the case of the 3D network structure, due to the extra dimension, 
the number of elements in parallel corresponds to the square of the number of the elements in series.
After counting all the contributions at the $n^\text{th}$ iteration, we have:
\begin{equation}
J^{3\text{D}}(t)=\frac{N_L-K_L}{(N_L+1)^2}\sum_{i=1}^{n-1}\frac{J_{l_i}(t)}{K_L^{i-1}}+\frac{N_L}{(N_L+1)^2}\frac{J_{l_n}(t)}{K_L^{n-1}}
\label{eq 3D}
\end{equation}

As for the 1D case, these sums may also be approximated by the integrals:
\begin{align}
J^{2\text D}(t)\sim \int_{l_1}^{l_n} J_{l}(t)\frac{di}{dl}dl\sim \int \tau^{\delta/\gamma-1}\left(1-e^{-t/\tau}\right)d\tau\\
J^{3\text D}(t)\sim  \int_{l_1}^{l_n}\frac{J_{l}(t)}{K_L^{i}} \frac{di}{dl}dl\sim \int \tau^{(\delta+\xi)/\gamma-1}\left(1-e^{-t/\tau}\right)d\tau
\label{eq 3D int}
\end{align}
The weak power law behaviours exponents become 
\begin{align}
\alpha^{2\text D}=\frac{\delta}{\gamma},\;\;\;\;\; \alpha^{3\text D}=\frac{\delta+\xi}{\gamma}
\end{align}

The 1D self-similar structure analysed here is simply a line of connected segments.
However, the 2D and 3D self-similar structures correspond to networks with non-integer fractal dimensions.
We may cover the whole structures (except a number of lines) with $(K_L^i)^D$ boxes ($D=2,3$ for the 2D or the 3D network, respectively)
of size $L/N_L^i$, yielding a fractal dimension $d_f=D\xi$ (if $d_f>1$).

The parameters $\delta$ and $\gamma$ depend on the choice of our particular model. 
Several possibilities may be considered.
It is reasonable to assume, as in Stokes's law, that $\eta_l$ is proportional to the viscosity of the solvent and to the size of the element, $\eta_l\sim l$.
Then, the retardation time $\tau_l=J_l \eta_l\sim l^{\delta+1}$, leading to $\gamma=\delta+1$.
If we take $J_l=1/G_l$ constant (this choice was done for the rheological stiffnesses of the SGM model, or the Rouse model \cite{rubinstein2003polymers};
in the latter case, it followed from the equipartition theorem),  $\delta=0$, and we may obtain the interesting result
\begin{align}
\alpha^{3\text D}=\frac{d_f}{3},\;\;\;\;\; (\delta=0, \;\;\gamma=\delta+1=1)
\end{align}
There are other reasonable scaling laws. We may for instance invoke the idea of springs in series to justify $J_l\sim l$.
In this case, $\tau_l\sim l^2$, and $\alpha^{3\text D}=(1+d_f/3)/2$. 

In this article, we have presented a new paradigm for soft solid or gelled materials.
We have shown that a self similar network, with a power law distribution of segment lengths, may lead to a 
generalised Kelvin-Voigt model with a power law spectrum of retardation times, $\cal L\sim\tau^{\alpha}$,
where $\alpha$ is related to the fractal dimension of the network.
We recover the weak power law behaviours $G'\sim G''\sim w^\alpha$ for $0<\alpha<1$, often observed in the cell cytoskeleton, and other soft materials.
For $\alpha<0$ or $\alpha>1$, we also obtain other interesting power law behaviours, which are characteristic of this effective generalised Kelvin-Voigt model.

The system here presented compares with the ``Soft Glassy Material''(SGM) model which, in what regards its linear viscoelastic regime, may be associated with 
a generalised Maxwell model with a power law spectrum of relaxation times ${\cal H} \sim1/\tau^{\alpha}$ (see \cite{patricio2015weak} for a detailed comparison).
With this model, we also recover the weak power law behaviours $G'\sim G''\sim w^\alpha$ for $0<\alpha<1$, but we have other, substantially different power law 
behaviours for $\alpha<0$ or $\alpha>1$ \cite{sollich1998rheological,patricio2015weak}. 
The generalised Maxwell model (and the SGM model) reflects a more fluidic system, which contrasts with the solid like or gelled generalised Kelvin-Voigt model 
(obtained from a self-similar network) described in this article.

Soft materials, including the cell cytoskeleton, are usually very complex materials,
in which we probably have a mixture of microscopic Kelvin-Voigt elements associated with Maxwell relaxation structures, both in series and in parallel,
leading to different scaling behaviours in different ranges of angular frequencies. Furthermore, the cytoskeleton is an active structure from
which we may expect novel behaviours, at least for particular ranges of characteristic times.
We hope, however, that this new solid like or gelled paradigm model may bring improved understanding of these rheological weak power law behaviours,
that appear so often in so many complex soft materials.


\bibliography{SGM}

\end{document}